\documentclass[10pt]{article}
\usepackage[latin1]{inputenc}
\usepackage{opex3}
\providecommand{\citeonline}[1]{\cite{#1}}

\usepackage{graphicx}
\usepackage{amsmath,amssymb,xspace}
\usepackage{textcomp} 

\DeclareMathAlphabet {\bmath}{OT1}{ptm}{b}{it}
\renewcommand{\boldsymbol}[1]{\bmath{#1}}

\newcommand{\mykeywords}{%
  wavefront sensing (010.7350); %
  adaptive optics (010.1080); %
  phase retrieval (100.5070); %
  atmospheric turbulence (010.1330); %
  inverse problems (100.3190); %
  telescopes (110.6770).}

\usepackage[bookmarksnumbered,bookmarksopen=true,pdfusetitle,%
             colorlinks=true,linkcolor=black,citecolor=black]{hyperref}%
\hypersetup{%
  pdfsubject={Optics Express 2008 paper by L. Mugnier et al. on Long-Exposure
              Phase Diversity},%
  pdfkeywords={\mykeywords}}

 \usepackage[firsttwo,bottomafter,dark,outline]{draftcopy}
 \draftcopyName{To be published in Optics Express}{60}

\newcommand{\NAOS}{\textsc{Naos}\xspace}
\newcommand{\RMS}{\textsc{rms}\xspace}
\newcommand{\SAXO}{\textsc{Saxo}\xspace}
\newcommand{\SPHERE}{\textsc{Sphere}\xspace}

\newcommand{\cf}{\emph{cf.}\xspace}
\newcommand{\etal}{~\emph{et al.}\xspace}
\newcommand{\ie}{\emph{i.e.},\xspace}
\newcommand{\eg}{\emph{e.g.},\xspace}

\newcommand{\htilde}[1][]{\tilde{h}_{\text{#1}}}
\newcommand{\h}[1][]{{h}_{\text{#1}}}
\newcommand{\mathopsf}{\langle \h[o]\rangle}

\newcommand{\bff}{\ensuremath{\mathbf f}\xspace}
\newcommand{\bfh}[1][]{\boldsymbol{h}_{\text{#1}}}

\newcommand{\bfi}[1][]{\boldsymbol{i}_{\text{#1}}}
\newcommand{\bfn}{\boldsymbol{n}}
\newcommand{\bfo}{\boldsymbol{o}}
\newcommand{\bfphi}{\boldsymbol{\phi}}
\newcommand{\mum}{\,\ensuremath{\text{\textmu{}m}}\xspace}
\newcommand{\cm}{\,\text{cm}\xspace}
\newcommand{\mm}{\,\text{mm}\xspace}
\newcommand{\nm}{\,\text{nm}\xspace}

\newcommand{\Times}{\!\times\!}

\newcommand{\norm}[1]{\| {#1} \|}

\graphicspath{{./}{./figures/}}

\begin{document}

\title{On-Line Long-Exposure Phase Diversity: a Powerful Tool for Sensing
  Quasi-Static Aberrations of Extreme Adaptive Optics Imaging Systems.}

\author{L. M. Mugnier, J.-F. Sauvage, T. Fusco, A. Cornia and S. Dandy}

\address{Office National d'\'{E}tudes et de Recherches A\'{e}rospatiales,
Optics Department, BP 72, F-92322 Ch\^{a}tillon cedex, France}

\email{mugnier@onera.fr}


\begin{abstract}
The phase diversity technique is a useful tool to measure and pre-compensate
for quasi-static aberrations, in particular non-common path aberrations, in
an adaptive optics corrected imaging system. In this paper, we propose and
validate by simulations an extension of the phase diversity technique that
uses long exposure adaptive optics corrected images for sensing quasi-static
aberrations during the scientific observation, in particular for
high-contrast imaging. The principle of the method is that, for a
sufficiently long exposure time, the residual turbulence is averaged into a
convolutive component of the image and that phase diversity estimates the
sole static aberrations of interest.

The advantages of such a procedure, compared to the processing of
short-exposure image pairs, are that the separation between static aberrations
and turbulence-induced ones is performed by the long-exposure itself and not
numerically, that only one image pair must be processed, that the estimation
benefits from the high SNR of long-exposure images, and that only the static
aberrations of interest are to be estimated. 
Long-exposure phase diversity can also be used as a phasing sensor for a
segmented aperture telescope. Thus, it may be particularly useful for future
planet finder projects such as EPICS on the European ELT.
\end{abstract}

\ocis{\mykeywords}





  \providecommand{\conferencedatename}{Date conf\'erence~: }
  \renewcommand{\conferencedatename}{Conference date: }

\section{Introduction}

Calibrating the quasi-static aberrations of a ground-based Adaptive Optics
(AO)~\cite{Roddier-l-99} corrected imaging system is an important issue,
especially for extreme AO high contrast instruments such as the proposed
planet finder instruments for the ESO and Gemini 8-meter telescopes.

Measuring these aberrations off-line, \ie during a day-time calibration on an
internal reference source, has been successfully applied to existing systems
such as \textsc{Naos}~\cite{Blanc-a-03a,Hartung-a-03} or Keck, and recently
refined for the \textsc{Sphere} project in order to achieve a nanometric
accuracy~\cite{Sauvage-a-07}. The main limitations of such a procedure directly
stem from its off-line nature: the aberrations located before the internal
reference source are not sensed, and the aberrations may evolve between the
day-time calibration and the night-time observation. These two limitations can
be circumvented by an appropriate calibration performed on-line, \ie during
night-time observations, as proposed in the following.

Two problems must be addressed in order to calibrate on-line the quasi-static
aberrations of the optical system made of the telescope, its AO system and the
camera: the first one is to distinguish between the turbulence-induced
component of the wavefront and the static one, which is the only wavefront of
interest for the problem at hand, and the second one is to sense all the
aberrations from the (potentially segmented) primary down to the focal plane
of the camera.

A wave-front sensor (WFS) is able to measure the aberrations seen by the
telescope on-line, but these consist of the sum of a turbulence-induced
component, which is partially compensated for by the AO, and a static
component.
Because the turbulence, whether corrected or not, evolves quickly, the WFS
measurements are generally performed with integration times that freeze the
turbulence evolution, typically a few milli-seconds.

There is today a large number of WFSs, which are thoroughly reviewed
in Ref.~\citeonline{Rousset-l-99a} and can be classified into two families:
pupil-plane sensors, such as the Hartmann-Shack and the curvature sensors, and
focal-plane sensors.
A focal-plane WFS is the only way to be sensitive to \emph{all} aberrations
down to the focal plane, and in particular to the so-called non-common path
aberrations of an AO system, which motivates our choice for a focal-plane WFS
in the following.
Estimating aberrations from a single focal-plane image of a point source is a
difficult problem known as phase retrieval. 
Phase-retrieval has two major limitations. Firstly, it only works with a point
source. Secondly, there is generally a sign ambiguity in the recovered phase,
i.e., the solution is not unique. Gonsalves~\cite{Gonsalves-a-82} showed that
by using a second image with an additional known phase variation with respect
to the first image such as defocus, it is possible to estimate the unknown
phase even when the object is extended and unknown. The presence of this
second image additionally removes the above-mentioned sign ambiguity of the
solution. This technique, referred to as phase diversity, has been
significantly developed in the past twenty years, both for wave-front sensing
and for imaging; see for instance
Refs.~\citeonline{Paxman-88,Paxman-92a,Lofdahl-a-94,Lee-a-97a,Thelen-a-99,Blanc-a-03b,Hartung-a-03},
and Ref.~\citeonline{Mugnier-l-06a} for a review.

The estimation of static aberrations from a series of short-exposure
phase-diversity data has been performed using a series of image pairs of
an astronomical object~\cite{Acton-a-96,Baba-a-01}. 
Lee\etal~\cite{Lee-a-97b}
have performed such a calibration of static aberrations with a series of
images instead of image pairs, and an original diversity: no additional
defocused image was used, and successive changes to the adaptive optics
introduced the required diversity.

In both approaches, the static aberrations are obtained as an empirical
average of the phase estimates corresponding to each short-exposure data.
This is notably suboptimal for at least three reasons:
\begin{itemize}
\item the images correspond to short integration times, and are consequently
  noisier than the corresponding long-exposure image pair, so each phase
  estimate suffers from this noise;
\item the computational cost is high because many short-exposure images
  must be processed in order to estimate the sought static aberrations;
\item the phase estimation accuracy may be penalized because the estimation
  must be performed on a number of phase parameters that is large enough to
  describe the short-exposure phase, whereas only a smaller number of these
  parameters may be of interest, if the sought static aberrations are of lower
  order than turbulence-induced ones;

\end{itemize}

In this paper, we propose and validate an extension of the phase diversity
technique that uses long-exposure AO-corrected images for sensing quasi-static
aberrations. This way, (1) the separation between quasi-static aberrations and
turbulence-induced ones is performed by the long-exposure itself and not
numerically, (2) only one image pair must be processed, (3) the estimation
benefits from the high SNR of long-exposure images, and (4) only the static
aberrations of interest are to be estimated.

\section{Principle of long-exposure phase diversity}

We consider a ground-based telescope observing Space through the turbulent
atmosphere. 
The long-exposure optical transfer function (OTF) of the atmosphere+instrument
system is the product of the OTF of the sole instrument $\htilde[s]$, called
static OTF in the following, by the atmospheric transfer function (ATF)
$\htilde[a]$~\cite{Roddier-81}:

\begin{equation}
  \label{eq-longexp-otf}
  \langle\htilde[o]\rangle = \htilde[s] \: \htilde[a] .
\end{equation}
The static OTF is a function of the unknown static aberrations, which are
coded in the phase function $\varphi$ in the aperture; let
$P$ be the indicator function of the aperture, \ie 1 in the aperture and 0
outside, $\htilde[s]$ is given by:

\begin{equation}\label{eq-static-otf}
  \htilde[s](\varphi) = P e^{i\varphi} \otimes  P e^{i\varphi}
\end{equation}
where $\otimes$ denotes auto-correlation.
The phase function $\varphi(u,v)$ is expanded on a basis $\{b_k\}$, which is
typically either Zernike polynomials or the pixel indicator functions in the
aperture:   $\varphi(u,v) = \sum_k \phi_k \: b_k(u,v)$,
where the summation is, in practice, limited to the number of coefficients
considered sufficient to correctly describe the static aberrations to be
estimated. We shall denote by $\bfphi$ the vector concatenating the set of
unknown aberration coefficients $\phi_k$.

Assuming phase perturbations with Gaussian statistics, the ATF at any spatial
frequency $\bff$ is given
by~\cite{Roddier-81}: 

\begin{equation}
  \label{eq-atf}
  \htilde[a](\bff) = e^{-\frac{1}{2} D_\phi (\lambda \bff)}
\end{equation}
where $\lambda$ is the imaging wavelength and $D_\phi$ is the phase structure
function.
If the turbulence is partially compensated by an AO system,
Equations~(\ref{eq-longexp-otf}) and~(\ref{eq-atf}) remain
valid~\cite{Roddier-l-99}, although slightly approximate because
the residual phase after AO correction is not stationary.

With these equations in hand, and assuming that the image is not larger than
the isoplanatic patch, we can now model the long-exposure image.
The image is recorded by a detector such as a CCD camera, which integrates the
flux on a grid of pixels. This can be conveniently modeled as the convolution
by a detector PSF $\h[d]$, assumed to be known in the sequel, followed by a
sampling operation. Using Eq.~(\ref{eq-longexp-otf}), the global long-exposure
PSF of the instrument is thus:
\begin{equation}
  \label{eq-global-psf}
  \h[le] = \h[d] \star \mathopsf   = \h[d]  \star \h[s] \star \h[a],
\end{equation}
where $\h[s]$ is the PSF due to static aberrations, given by the inverse Fourier
transform of Eq~(\ref{eq-static-otf}), $\h[a]$ is the atmospheric PSF
given by the inverse Fourier
transform of Eq.~(\ref{eq-atf}), and $\star$ denotes convolution. 

Due to the inevitable sampling and noise of the detection processes, the image
$\bfi[f]$ recorded in the focal plane is the noisy sampled convolution of the
long-exposure point-spread function (PSF) $\h[le]$ with the observed object
$o$.
%
%
This model is generally approximated by a noisy discrete convolution with the
sampled version $\bfo$ of the object $o$: 

\begin{equation}
\label{eq-image}
\bfi[f] = \bfh[le] \star \bfo + \bfn , 
\end{equation}
where $\bfh[le]$ is the sampled version of $\h[le]$, and $\bfn$ is a corruptive
noise process. If the noise is not additive and independent from the
  noiseless image, for instance if it is predominantly photon noise, then
  Eq.~(\ref{eq-image}) should read $\bfi = \bfh[le] \star \bfo \diamond \bfn$,
  with the symbol $\diamond$ representing a pixel-by-pixel
  operation~\cite{Demoment-89}. For legibility we shall keep the additive
  notation.

Combining Eq.~(\ref{eq-image}) with the discrete counterpart of
Eq.~(\ref{eq-global-psf}) yields the following discrete image model for the
focused and defocused images respectively:
\begin{eqnarray}
\label{eq-image2}
\bfi[f] &=& \bfh[d]  \star \bfh[s]^{(\bfphi)} \star \left( \bfh[a] \star \bfo
\right) + \bfn \\
\label{eq-image-defoc}
\bfi[d] &=& \bfh[d]  \star \bfh[s]^{(\bfphi+\bfphi_d)} \star \left( \bfh[a] \star \bfo
\right) + \bfn' ,
\end{eqnarray}
where $\bfh[d]$, $\bfh[s]$ and $\bfh[a]$ are the sampled versions
of $\h[d]$, $\h[s]$ and $\h[a]$, $\bfphi_d$ is the known additional phase
introduced in image $\bfi[d]$, and the superscripts on $\bfh[s]$ are reminders
of the aberrations that enter the static PSF of each image.

Let $\bfo'$ be the convolution of the atmospheric PSF with the
observed object:

\begin{equation}
\bfo'=\bfh[a] \star \bfo
\label{eq-oprime}
\end{equation}

The phase-diversity data model of Eqs.~(\ref{eq-image2})
and~(\ref{eq-image-defoc}) is strictly identical to the one that would be
obtained by imaging the pseudo-object $\bfo'$ in the absence of turbulence and
with the same static aberrations.
Thus, all the phase estimation methods developed for short-exposure images, in
which the OTF of the system is completely described by a phase function, can
be applied here to estimate the sole static aberrations. 

Additionally, it is well-known that, for a given noise level, the estimation
quality of the aberrations in phase diversity depends on the spectral content
of the observed scene. Thus, in the method proposed here, the estimation
quality of the aberrations will depend both on the spectral content of the
observed object and on the ATF, \ie on the
turbulence correction quality provided by the AO.

  The appropriate implementation of this long-exposure phase diversity
  technique depends on the type of instrument. For non coronagraphic
  instruments, one should use images provided by the science sensor; the
  defocused image can be either obtained simultaneously with the focused
  science image by means, \eg of a beamsplitter, or alternately. In the
  latter case, the deformable mirror itself can be used to provide the
  defocus~\cite{Blanc-a-03a,Hartung-a-03}. In both cases the fraction of the
  incoming flux allotted to the defocused image may be notably less than 50\%,
  in order to maximize the flux on the scientific data. Indeed, if the
  quasi-static aberrations are measured at intervals of, \eg a half-hour,
  defocused images must only be available with that kind of rate.

  For coronagraphic systems, for which the aberrations to be minimized are the
  ones located before the coronagraph, one may use a beam-splitter and an
  auxiliary image sensor located just before the coronagraph. For the \SPHERE
  instrument such a sensor actually already exists in the design for centering
  the star image on the coronagraph: it is the so-called differential tip-tilt
  sensor. This sensor could easily be adapted and used for the long-exposure
  phase diversity measurements. As in the non coronagraphic case, the
  defocused image may be obtained simultaneously with the focused image by use
  of a beamsplitter, or alternately by means of a
  longitudinal displacement of the sensor by a few millimeters.

\section{Chosen phase estimation method}

Among all the possible estimation methods (see, \eg
Ref.~\citeonline{Mugnier-l-06a} for a review) in this paper we choose, for
simplicity, the conventional least-squares joint estimation of the phase
$\bfphi$ and the object $\bfo'$, with a regularization on both quantities:
$(\hat{\bfo}', \hat{\bfphi}) = \arg\min J(\bfo', \bfphi) $ with

\begin{equation}
  \label{eq-criterion}
  J(\bfo', \bfphi) = 
\frac{1}{2\sigma_n^2} 
\norm{\bfi[f] - \bfh[d]  \star \bfh[s]^{(\bfphi)} \star \bfo' }
+
\frac{1}{2\sigma_{n'}^2}
\norm{\bfi[d] - \bfh[d]  \star \bfh[s]^{(\bfphi+\bfphi_d)} \star \bfo' } 
+R_o(\bfo') + R_\phi(\bfphi),
\end{equation}
where $\sigma_n^2$ and $\sigma_{n'}^2$ are the noise variances of the two
images, estimated beforehand. 
The object regularization is chosen as a quadratic function, so that the whole 
criterion $J$ is quadratic with respect to $\bfo'$ and thus has an unique,
closed-form solution $\hat{\bfo'}(\bfphi)$ for a given phase $\bfphi$. 
This allows one to replace the optimization of $J(\bfo', \bfphi)$ with that of
criterion $J'(\bfphi) \triangleq J(\hat{\bfo'}(\bfphi),\bfphi)$, as commonly
done in the unregularized case~\cite{Gonsalves-a-82,Paxman-92a}.

Following the findings of Blanc~\cite{Blanc-t-02,Blanc-a-03b}, we
under-regularize the object
in order to best estimate the phase. This strategy is supported by the fact
that it yields a phase estimation with satisfactory asymptotic properties, as
shown in Ref.~\citeonline{Idier-a-05}, and that these properties hold even if
the noise is not Gaussian.

  Concerning the phase, we choose the basis of the pixel indicator functions
  rather than, \eg a truncated basis of Zernike polynomials, in order to model
  and reconstruct phases with a high spatial frequency content. Because of the
  potentially large number of phase unknowns we are lead to regularize the
  phase estimation. To this aim, we use a functional proposed specifically for
  such a phase basis in Refs.~\cite[chap.~7]{Blanc-t-02}
  and~\cite[Sect.~8]{Mugnier-l-06a}, which is recalled below:

\begin{align}
R_\phi(\bfphi)=\sum_{(l,m) \in
S}&\left[\left|e^{j(\phi_{l-1,m}-\phi_{l,m})}-e^{j(\phi_{l,m}-\phi_{l+1,m})}\right|^2 \right.
+\left.
  \left|e^{j(\phi_{l,m-1}-\phi_{l,m})}-e^{j(\phi_{l,m}-\phi_{l,m+1})}\right|^2
\right] ,\nonumber
\end{align}
where the summation is done on all the pixels within the pupil ($S$ is the pupil
support). Furthermore, we impose a strict support constraint \ie all terms
$|...|^2$ that contain, at least, a pixel out of the pupil support, are
suppressed.

This regularization function has been constructed in such a way that it is
insensitive, as the data is, to a global piston, to tip-tilt, and to any
$2\pi$ variation of the phase on any pixel. This way, no local minimum is
introduced by the regularization into the minimized criterion.

\section{Validation by simulations }

We shall now validate the proposed method by simulations. We shall essentially
study the influence of the exposure time. Indeed, the main specific assumption
of the proposed method lies in Eq.~(\ref{eq-longexp-otf}), because the
factorization of the OTF in a static OTF and an ATF is strictly valid only if
the turbulence is perfectly averaged \ie for an infinite exposure time. Note
that by exposure time we mean the (finite) number of independent turbulence
realizations, not the noise level. In all the simulations presented here, we
have considered noiseless images. We have checked that the behavior of the
phase estimation in the proposed method with respect to the noise level is not
specific and is the same as conventional phase diversity with short-exposure
images: the estimation error is usually proportional to the average standard
deviation of the noise in the images~\cite{Meynadier-a-99,Blanc-a-03b}.
Then we shall
briefly study the influence of the AO correction quality.

\subsection{Conditions of simulation}
\label{sec-conditions_simulation}

We consider here a point-like source, observed with an $8$\:m ground-based
telescope, equipped with AO. The simulations take into account both the
AO-corrected turbulence and static aberrations. The baseline adaptive optics
system considered is the high-performance AO system \SAXO~\cite{Fusco-a-06b}
of the \SPHERE~\cite{Beuzit-p-07-correct} instrument. We use a Fourier-based
simulation method that describes the AO via the spatial power spectrum of the
residual phase~\cite{Jolissaint-a-06} and is presented in
Ref.~\citeonline{rconan-p-04}. The simulation takes the following realistic
set of parameters: a $41\times41$ subaperture Shack-Hartmann, a $1.2\:$kHz
sampling frequency, a guide star of magnitude $8$, and a Paranal-like
turbulence profile, with a seeing of $0.8$\:arcsec at $0.5\mum$.
Static aberrations are randomly generated according to a $f^{-2}$ spectrum,
$f$ being the spatial frequency in the pupil, with a total wavefront error of
$0.23$~radian \RMS at $1.6\mum$, \ie $60$~nm \RMS. 

For some of the simulations, the number of actuators on a pupil diameter will
be decreased from $40\Times40$ to $14\Times14$ and $7\Times7$ in order to
study the influence of the AO correction quality. The above simulation
conditions lead to a phase variance of the residual turbulence which is
respectively $1$, $7$ and $21$ time(s) the variance of the static aberrations.

Two simultaneous long-exposure images are simulated, with a phase diversity
between these images consisting of a $1.814$~radian \RMS defocus. 
%
  The corresponding defocus distance is proportional to the square of the
  f-number of the system and, at $1.6\mum$, is $2.9\mm$  for an $f/15$ system
  such as \NAOS and $2\cm$ for an $f/40$ system such as \SPHERE.
These  images are simulated following two schemes:
%
%
\begin{itemize}
  \item Finite exposure time images are made of the summation of $N$
    short exposures. The long-exposure OTF is then 
    the sum of $N$ short-exposure OTFs, each of which being computed through
    Eq.~(\ref{eq-static-otf}), with a phase $\varphi$ composed of the sum
    of the static aberrations and of the instantaneous AO-corrected turbulent
    wavefront. The turbulent wavefront is randomly generated from a PSD that
    takes into account both the turbulence profile and the AO
    correction~\cite{rconan-p-04}.

  \item Infinite exposure time images are not computed  as an
    empirical average. Instead, the residual phase structure function $D_{\varphi}$ is
    computed from the PSD of the AO-corrected turbulent wavefront, then the
    ATF $\htilde[a]$ is computed via Eq.~(\ref{eq-atf}), and the images
    $\bfi[f]$ and $\bfi[d]$ are computed according to
    Equations~(\ref{eq-image2}) and~(\ref{eq-image-defoc}).
\end{itemize}

\subsection{Influence of exposure time}

\begin{figure}[!htb]
\begin{center}\leavevmode
  \includegraphics[width=0.75\linewidth]{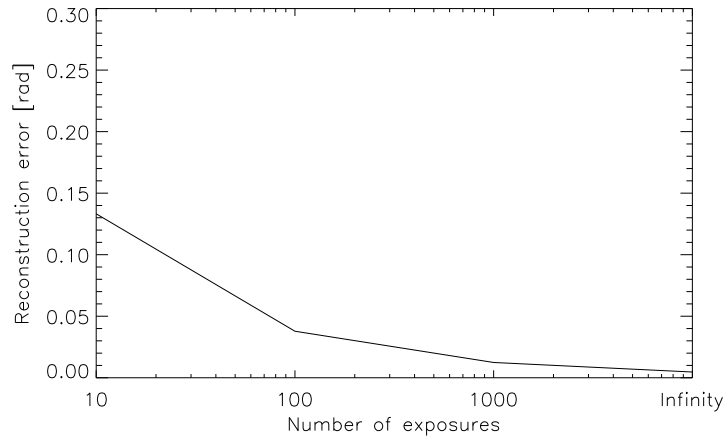}
  \caption{Evolution of the reconstruction error with the exposure time (in
    number of independent turbulence realizations). Static aberrations are
    randomly generated according to conditions described in
    Subsection~\ref{sec-conditions_simulation}. AO correction is assumed to be
    performed by a \SAXO-like system. }
  \label{fig-error_nbexpo_SPHERE}
  \end{center}
\end{figure}

We first consider the case of a high performance, \SAXO-like correction.
Figure~\ref{fig-error_nbexpo_SPHERE} shows the evolution of the reconstruction
error with the number of exposures used for the simulation. The first points
(from $10$ to $1000$ exposures) are simulated with a finite exposure time,
whereas the last point (noted infinity on the X-axis) is simulated with an
infinite exposure time. 
%
  Because the correlation time of corrected turbulence is typically $10$ to
  $100$ milliseconds depending on turbulence parameters and on the AO
  correction quality, the simulations with $1000$ exposures correspond to an
  integration time between $10$ and $100$ seconds.

The reconstruction error decreases with the number of exposures, down to very
weak values (less than $0.01$~radian) for an infinite exposure time. As the
number of exposures increases, Eq.~(\ref{eq-longexp-otf}) becomes 
more valid, turbulence residuals become better fitted by a
modification of the estimated object only (into an object $\bfo'=\bfh[a] \star
\bfo$, \cf Eq.~\ref{eq-oprime}), and the estimated phase is eventually only
the set of static aberrations. Incidentally, we have checked that for a single
turbulent exposure, the phase estimated is the sum of the turbulent wavefront
and the static aberrations.

This simulation study shows that the estimation of static aberrations from a
single pair of turbulence-degraded images is possible. The quality of
aberration reconstruction is directly linked to the convergence of the images
towards long-time exposures. For the AO system and the level of static
aberrations considered here, an integration time corresponding to a thousand
independent turbulence realizations yields a phase estimation error of about
$0.012$~radian, close to that obtained with an infinite exposure time.
  At $1.6\mum$, this number translates into a $3\nm$~\RMS optical path
  difference. This precision is compatible with the $5-10\nm$~\RMS static
  aberration residual that is needed for the detection of warm Jupiters on an
  $8$-meter telescope. Incidentally, we see that with less than $1000$
  exposures the required precision would not be obtained for such a mission.

\subsection{Influence of AO correction quality}

\begin{figure}[!htb]
\begin{center}\leavevmode
  \includegraphics[width=0.75\linewidth]{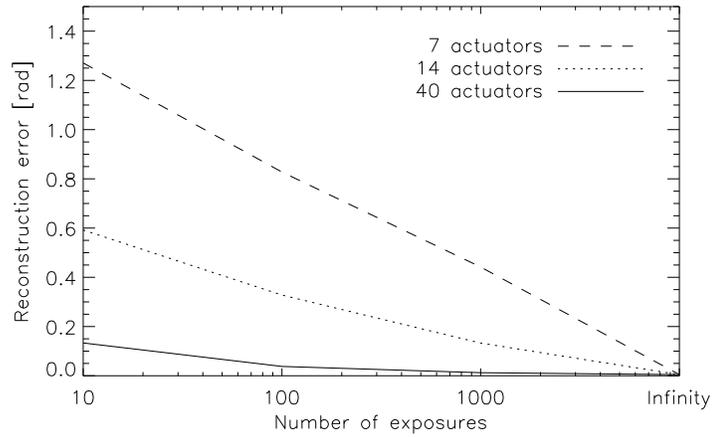}
  \caption{Evolution of the reconstruction error with the exposure time, for
    several levels of AO correction.
    Static aberrations are randomly generated according to conditions
    described in Subsection~\ref{sec-conditions_simulation}.
  \label{fig-error_nbexposures}
}
  \end{center}
\end{figure}

Figure~\ref{fig-error_nbexposures} shows the evolution of the estimation
error, for different correction qualities, obtained here simply by varying the
number of actuators of the AO system, all other parameters being equal. One
can see that as the correction quality degrades, the turbulence residuals are
more important and thus the estimation of the static aberrations needs more
exposures for the same error level. The estimation errors for an infinite
exposure time are equivalent for all three correction qualities.
%
%
%

  In the case of a \NAOS correction with $14\Times14$ actuators and $1000$
  exposures, the phase estimation error is about $0.13$~radian, which at
  $2.2\mum$ translates into a $45\nm$~\RMS optical path difference. This is
  almost three times smaller than the residual static aberrations of
  $120\nm$~\RMS measured on \NAOS-\textsc{Conica} \emph{after} off-line phase
  diversity measurement and correction~\cite{Hartung-a-03}. The on-line
  long-exposure phase diversity technique could thus be an attractive way to
  calibrate quasi-static aberrations on non-extreme AO systems too.

We now detail the spectral analysis of the estimated aberrations. In an AO
system, the number of actuators determines the highest spatial frequency of
the turbulence to be corrected. This parameter therefore directly impacts on
the spectral content of the turbulence residuals. On the following figures we
plot the circularly averaged spectra of the estimated aberations with respect
to the spatial frequency, for different AO correction levels.

\begin{figure}[!htb]
\begin{center}\leavevmode
  \includegraphics[width=0.75\linewidth]{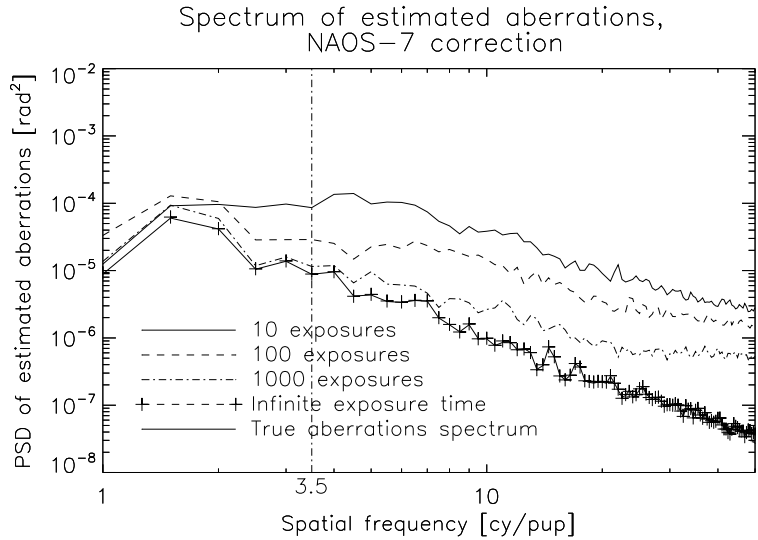}
  \caption{Spatial spectrum of the estimated aberrations, in the case of a
    \NAOS-7 correction. 
    The vertical line represents the maximum spatial frequency that is
    corrected by the AO system.}%
  \label{fig-spectres_prof_7act}
  \end{center}
\end{figure}

Figure~\ref{fig-spectres_prof_7act} shows the error spatial spectrum in the
case of a \NAOS correction ($7\Times7$ actuators), for different exposure time. One
can see that the spectrum of estimated aberrations get closer to that of the
true ones as the number of exposures increases, and that the convergence is
slower for the uncorrected frequencies of the turbulence (above
$3.5$\:cycles/pupil for this $7\Times7$ actuator system).

\begin{figure}[!htb]
\begin{center}\leavevmode
  \includegraphics[width=0.75\linewidth]{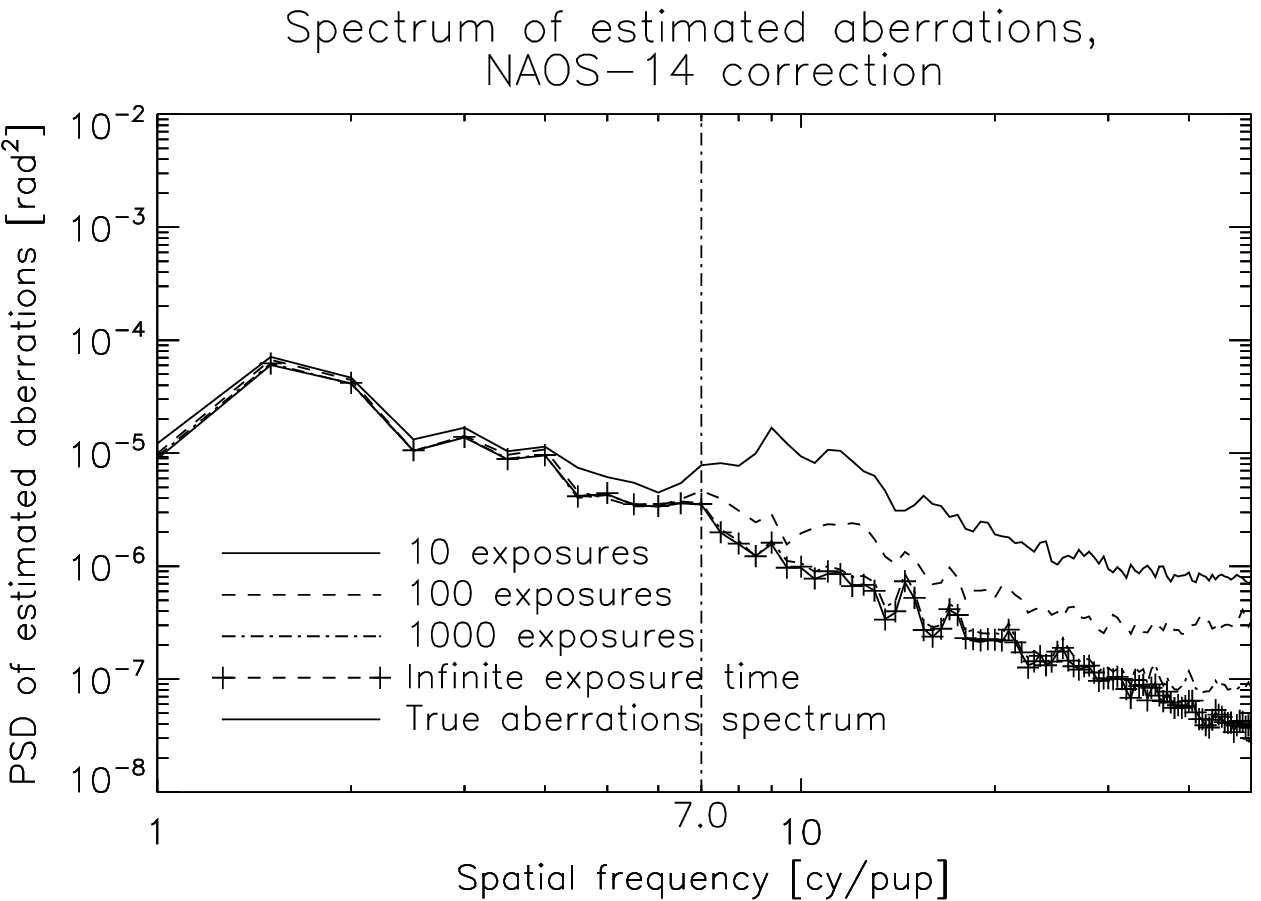}
  \caption{Spatial spectrum of the estimated aberrations, in the case of a
    \NAOS-14 correction. 
    The vertical line represents the maximum spatial frequency that is
    corrected by the AO system.}%
  \label{fig-spectres_prof_14act}
  \end{center}
\end{figure}

\begin{figure}[!htb]
\begin{center}\leavevmode
  \includegraphics[width=0.75\linewidth]{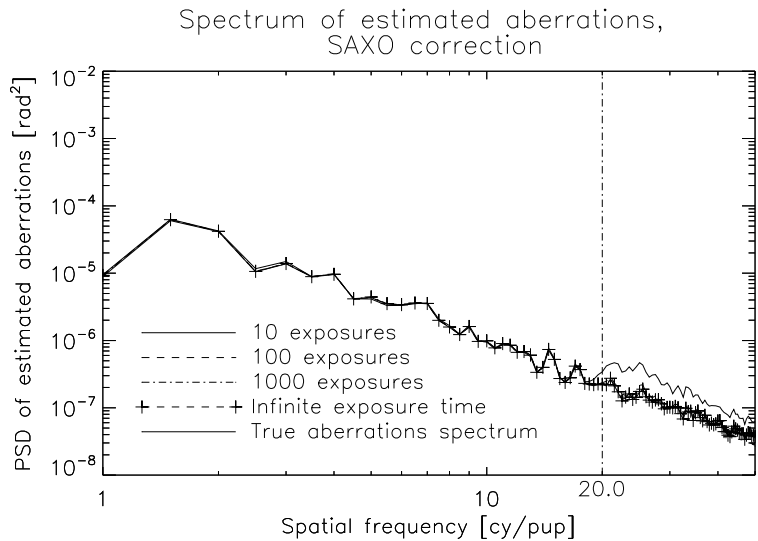}
  \caption{Spatial spectrum of the estimated aberrations, in the case of a
    \SAXO/\SPHERE correction. 
    The vertical line represents the maximum spatial frequency that is
    corrected by the AO system.}%
  \label{fig-spectres_prof_40act}
  \end{center}
\end{figure}

In the case of a higher correction (respectively $14\Times14$ and $40\Times40$
actuators, on Figures~\ref{fig-spectres_prof_14act}
and~\ref{fig-spectres_prof_40act}), the turbulence residuals are reduced and
the convergence of the spectrum to true spectrum is consequently faster.
Moreover, as in the \NAOS-7 case, convergence is notably faster in the
corrected part of the aberration spectrum, which correspond to a limit of $7$
and $20$\:cycles/pupil respectively.

\section{Conclusion and perspectives}

The phase diversity technique is a powerful tool to measure and pre-compensate
for quasi-static aberrations, in particular non-common path aberrations, in an
AO-corrected imaging system.

So far it has, to the best of our knowledge, either been used off-line on an
internal calibration source, or on-line from short-exposure
turbulence-degraded data. Each approach has its own limitations, discussed in
this paper.
We have proposed and validated by simulations an extension of the phase
diversity technique that uses long exposure AO corrected images for sensing
quasi-static aberrations.

The principle of the method is that, for a sufficiently long exposure time,
the residual turbulence is averaged into a convolutive component of the image
and that phase diversity estimates the sole static aberrations of interest.

Technically, the advantages of such a procedure, compared to the processing of
short-exposure image pairs, are that the separation between static aberrations
and turbulence-induced ones is performed by the long-exposure itself and not
numerically, that only one image pair must be processed, that the estimation
benefits from the high SNR of long-exposure images, and that only the static
aberrations of interest are to be estimated. 
Compared to pupil-plane wavefront sensing techniques, phase diversity has the
advantages that it senses all aberrations down to the focal plane and that the
hardware is extremely simple.

From a system point of view, on-line long-exposure phase diversity opens a new
area of applications, in particular it will allow one to correct in real time
(meaning during the scientific exposure) for any evolution of instrumental
defects. This may be considered for improvements to the \SPHERE instrument and
should significantly improve the overall system detectivity.

This technique can also be used as a phasing sensor for a segmented aperture
telescope. Indeed, phase diversity can be applied to the peculiar aberrations
constituted by the differential tip-tilts and pistons of such a
telescope.

Thus, long-exposure phase diversity may be particularly useful for the future
Planet Finder project on the E-ELT called EPICS~\cite{Kasper-p-07-correct}.
Indeed, on the one hand, for this project, the detectivity requirements are by
far more stringent than for \SPHERE and the on-line correction of non-common path
aberrations is mandatory. And on the other hand, without any opto-mechanical
modification of the sensor, long-exposure phase diversity should enable very
accurate measurements of the segments' phasing.

Lastly, for some applications it may be useful to estimate not only the static
aberrations but also the ATF, which is part of the estimated quantities. The
method proposed herein can be refined to estimate the ATF precisely using (a)
a parametrization of the ATF through the phase structure function and (b) a
specific space-varying regularization criterion for the ATF along the lines
of~\cite{Sauvage-p-06,Sauvage-t-07}.

Short-term perspectives include a more thorough study of the performance of the
long-exposure phase diversity technique, coupled with a global system
analysis.

\end{document}